\documentstyle[12pt,fleqn]{article}
\def\beeq{\begin{equation}}
\def\eneq{\end{equation}}
\def\beeqa{\begin{eqnarray}}
\def\eneqa{\end{eqnarray}}

\addtocounter{section}{-1}
\begin{document}

\begin{center}

\vspace{2cm}

{\large {\bf {Long-range excitons in conjugated polymers\\
with ring torsions\\
} } }

\vspace{1cm}

{\rm Kikuo Harigaya\footnote[1]{E-mail address: 
\verb+harigaya@etl.go.jp+; URL: 
\verb+http://www.etl.go.jp/~harigaya/+}}

\vspace{1cm}

{\sl Physical Science Division,
Electrotechnical Laboratory,\\ 
Umezono 1-1-4, Tsukuba, Ibaraki 305, Japan}

\vspace{1cm}

(Received~~~~~~~~~~~~~~~~~~~~~~~~~~~~~~~~~~~)
\end{center}

\vspace{1cm}

\noindent
{\bf Abstract}\\
Ring torsion effects on optical excitation properties in
poly({\sl para}-phenylene) (PPP) and polyaniline (PAN) 
are investigated by the intermediate exciton formalism.  
Long-range excitons are characterized, and the long-range 
component of the oscillator strengths is calculated. 
We find that ring torsions affect the long-range excitons 
in PAN more easily than in PPP, due to the larger torsion 
angle of PAN and the large number of bonds whose hopping 
integrals are modulated by torsions.

\pagebreak

\section{Introduction}

Recently, we have been studying structures of photoexcited states 
in electroluminescent conjugated polymers: poly({\sl para}-phenylene) 
(PPP), poly({\sl para}-\-phen\-yl\-ene\-vinylene) (PPV), 
poly({\sl para}-phenylenedivinylene) (PPD), and so on [1-3].
We have introduced the viewpoint of long-range excitons in
order to characterize photoexcited states where an excited
electron-hole pair is separated over a single monomer of the 
polymer.  We have shown [1,2] that a long-range exciton feature 
starts at the energy in the higher energy side of the lowest 
feature of the optical absorption of PPV.  The presence of 
the photoexcited states with large exciton radius is essential 
in mechanisms of the strong photocurrents observed in this 
polymer.  In ref. [3], we have compared properties of excitons 
in PPV-related polymers.  The  oscillator strengths of the 
long-range excitons in PPP are smaller than in PPV, and those 
of PPD are larger than in PPV.  Such relative variation is 
due to the difference of the number of vinylene bonds.

It is known that the PPV chain is nearly planer and the phenyl
rings are not distorted each other, as observed in X-ray 
analysis [4].  However, ring torsions, where even number phenyl 
rings are rotated in the right direction around the polymer axis 
and odd number phenyl rings are rotated in the left direction,
have been observed in PPP [5].  The polymer structure of PPP is 
shown schematically in Fig. 1 (a).  The ring torsions originate 
from the steric repulsion between phenyl rings, because the 
distance between the neighboring rings in PPP is smaller than 
that of PPV.  In a simple tight binding model, the ring torsion 
modulates the nearest neighbor hopping integral $t$ as
$t {\rm cos}\Psi$, where $\Psi$ is the torsion angle.  If the 
angle $\Psi$ is sufficiently large, the motion of electrons 
between phenyl rings would be hindered and also the exciton 
radius of photoexcited states would become shorter.  The 
contribution from long-range excitons will be smaller from 
that of the calculations [3] where the planer structure of 
PPP has been assumed.  The actual torsion angle about 23$^\circ$ [5]
may influence the results of the previous calculations.
The first purpose of this paper is to examine ring torsion
effects on optical excitations of PPP.  We will show that
the magnitude of the torsion $\Psi \sim 23^\circ$ does not
change the component of the long-range excitons so much.

The another example of polymers where ring torsions are present
is polyaniline (PAN) (leucoemeraldine base).  The schematic 
structure is displayed in Fig. 1 (b).  The phenyl rings and NH 
units are arrayed alternatively in the chain direction.  The 
magnitude of the torsion of phenyl rings in PAN is about 
56$^\circ$ [6].  This is larger than that of PPP, and thus the 
photoexcited states in PAN will be influenced more strongly by the 
torsion.  In the last half of this paper, we shall look at this 
problem.  We will show that the long-range component of the 
oscillator strengths at $\Psi = 56^\circ$ is about half 
of the magnitude of the system without ring torsions. 
Long-range excitons in PAN are hindered by ring torsions
more easily than those in PPP.  This is due to the fact that
the torsion angle is larger in PAN, and also that two bonds, 
whose hopping integrals are modulated as $t {\rm cos}\Psi$,
are present between neighboring phenyl rings in PAN while one
such bond is present between phenyls in PPP.

This paper is organized as follows.  In the next section, 
ring torsion effects on photoexcited states in PPP are 
reported.  The results of PAN are given in \S 3, and 
the paper is concluded with a summary in \S 4.

\section{Ring torsion effects in PPP}

The model used in [3] is modified in order to take account
of ring torsions.  The Hamiltonian in [3] is composed of
the tight binding part and the term of long-range Coulomb 
interactions among electrons.  In the tight-binding part,
the nearest neighbor hopping integral is modulated linearly
with respect to the bond length change by the electron-phonon 
coupling $\alpha$.  In the term of Coulomb interactions, the
parametrized Ohno potential is used among all the carbon
sites in order to describe exciton effects.  The form of
the Ohno potential is $1/\sqrt{(1/U)^2 + (r/a V)^2}$, where
$U$ and $V$ are the strengths of the onsite and long-range 
interactions.  We have assumed the following parameters in [3]: 
the mean bond length $a = 1.4$\AA~, $\alpha = 2.59t$/\AA, 
the harmonic spring constant $K=26.6t$/\AA$^2$, $U=2.5t$, 
and $V=1.3t$ ($t$ is the nearest neighbor hopping integral
of the system with the equal bond length $a$).

In this section, the torsion angle between neighboring phenyl 
rings $\Psi$ is taken into account in the hopping integral 
without ring torsions as $t-\alpha y \Rightarrow (t-\alpha y) 
{\rm cos} \Psi$ in the model reviewed above.  Here, $y$ is the 
length change of the corresponding bond.  The model is solved 
by the mean field approximation and electron-hole excitations 
are calculated by the intermediate exciton formalism.  The 
long-range component of photoexcited states are characterized 
as we have done in the previous paper [3].

Figure 2 shows the optical absorption spectra calculated
for the system with the torsion angle $\Psi = 23^\circ$ 
[5] and the monomer number $N_{\rm m}=20$.  The spectral 
shapes are nearly independent of the monomer number at 
$N_{\rm m}=20$, as reported in [3].  The polymer without
ring torsions with the open boundary is in the $x$-$z$ plane.  
The electric field of light is parallel to the chain and in 
the direction of the $x$-axis in Fig. 2 (a), and it is 
perpendicular to the chain and is along with the $y$-axis in 
Fig. 2 (b).  The electric field is along with the $z$-axis
in Fig. 2 (c).  The orientationally averaged spectra with 
respect to the electric field are shown in Fig. 2 (d).
In each figure, the bold line shows the total absorption
and the thin line shows the contribution from the long-range
excitons where photoexcited electron and hole are separated
over more than the spatial extent of the single monomer.
In Fig. 2 (a), there are two main features around 1.4$t$
and 2.4$t$, where quantities with energy dimensions are expressed 
in units of $t$.  There is a feature of long-range excitons 
around $\sim 2.0t$ at the higher energy side of the 1.4$t$ main 
peak.  In contrast, the 2.4$t$ feature does not have so strong 
long-range component due to the almost localized nature of 
excitons.  This is similar to the calculations in [3], and the 
ring torsions do not change the almost localized character of 
the feature around 2.4$t$ [7].  Figure 2 (b) shows the case 
where the electric field is parallel to the $y$-axis.  If the 
ring torsions are not present, the oscillator strengths are 
zero in this case.  However, they are finite and the maximum
of the bold line in Fig. 2 (b) is one order of magnitudes
smaller than that of Fig. 2 (a).  Two main features of the 
energies of 2.2$t$ and 2.8$t$ are derived from those of the 
case with the electric field in the $z$-direction, shown in 
Fig. 2 (c).  These features in Fig. 2 (c) contribute dominantly, 
and their oscillator strengths are of the same order as those 
of Fig. 2 (a).  There are long-range components of these 
two features around the energies about 2.5$t$ and 3.3$t$.  
Figure 2 (d) shows the optical spectra where the electric 
field is orientationally averaged.  The overall spectral 
shape is similar to that in the case without ring torsions 
$\Psi=0$ [3].  However, the spectral width decreases slightly
due to the smaller hopping interactions between neighboring
phenyl rings.

Next, we vary the torsion angle $\Psi$ in the Hamiltonian,
and look at changes in optical spectra.  We calculate the
long-range component in the total oscillator strengths, as
we have done in [1-3].  We search for its variations as a
function of $\Psi$.  Figure 3 shows the calculated results.
The closed squares are the data of the orientationally 
averaged absorption.  The open circles, squares, and triangles 
indicate the data for the cases with the electric field 
parallel to the $x$-, $y$-, and $z$-axes, respectively.
We find that the average long-range component weakly depends 
on $\Psi$, while $\Psi$ is as large as about 40$^\circ$.
Thus, we have examined ring torsion effects on optical 
excitations of PPP.  We have found that the magnitude of 
the torsion $\Psi \sim 23^\circ$ does not change the component 
of the long-range excitons so much from that of the torsion
free system.  However, the long-range component suddenly 
deceases at larger $\Psi$, and becomes zero at $\Psi=90^\circ$.  
This is a natural conclusion of the broken conjugations due to
strong ring torsions.  The long-range components with the
field in the $y$- and $z$-directions are generally larger
than that of the $x$-direction case.  This property has been
seen in ref. [3], too.

\section{Ring torsion effects in PAN}

It is of some interests to investigate ring torsion effects 
in conjugated polymers where torsion angles are larger than 
that of PPP.  In this section, we look at torsion effects 
in PAN as a typical example of such the polymers.  The polymer 
structure is shown schematically in Fig. 1 (b).  The zigzag
geometry of PAN is taken into account in the actual 
calculations, even though it is not shown in Fig. 1 (b).  For 
model parameters, we have assumed the negative site energy
$E_{\rm N} = -0.571t$ at the N sites, the hopping integral
$t_{\rm C-N} = 0.8t$ between carbon and nitrogen, the onsite 
interaction strength $U=2t$, and the magnitude of long-range 
interactions $V=1t$.  The parameters of $E_{\rm N}$ and 
$t_{\rm C-N}$ are taken from [8] as representative values.
We show the results with the monomer number $N_{\rm m}=20$.
We have taken several combinations of Coulomb interaction 
parameters and have looked at changes of results.  We have
found that the the long-range component as a function of $\Psi$
depends weakly on $U$ and $V$.  Thus, we report the results for
one combination of Coulomb parameters.

Figure 4 show the calculated optical absorption spectra of PAN
with the torsion angle $\Psi = 56^\circ$.  The three dimensional
coordinates of the polymer are similar to those of PPP in Fig. 2.
In Fig. 4 (a), the electric field is along with the polymer axis,
and is in the $x$-direction.  There are two main features around
0.6$t$ and 2.6$t$.  The former has smaller long-range exciton
feature than the latter.  As we observe in the band structure
of PAN [8], the lowest optical excitation among the band gap 
has a certain magnitude of dispersions, and thus its long-range 
component might be observed.  However, this lowest exciton is
dipole forbidden in the system without ring torsions when the 
electric field is parallel to the chain direction.  There is 
the second lowest optical excitation, which is almost localized
and is dipole allowed in the system without torsions, at the excitation
energy near the lowest exciton.  Thus, the long-range component
of the 0.6$t$ feature is suppressed.  Figure 4 (b) shows the 
weak optical absorption originated from the finite torsions
as we have found for the PPP in Fig. 2 (b).  Figure 4 (c) shows
the optical spectra when the electric field is in the $z$-direction.
Two features around 1.2$t$ and 2.6$t$ have certain magnitudes
of oscillator strengths due to long-range excitons.  This 
qualitative  property is similar to that in Fig. 2 (c).
Then, Fig. 4 (d) shows the optical spectra where orientational
average is performed.  We find excitation features superposed
from those of Figs. 4 (a), (b), and (c).

Finally, we show the long-range component of the oscillator
strength as a function of the torsion angle $\Psi$ in Fig. 5.
In contrast to PPP, the long-range component decreases smoothly
as $\Psi$ increases.  The long-range component of the total
oscillator strengths at the observed $\Psi = 56^\circ$ is about 
half of the magnitude of the system without ring torsions. 
Thus, the long-range excitons in PAN can be hindered by ring 
torsions more easily than those in PPP.  This is due to the fact 
that the torsion angle is larger in PAN, and also that two bonds, 
whose hopping integrals are modulated as $t {\rm cos}\Psi$,
are present between neighboring phenyl rings in PAN while one
such kind of bond is present between phenyls in PPP.

\section{Summary}

We have examined the ring torsion effects which might interrupt
delocalizations of excitons in the chain direction of the conjugated
polymers PPP and PAN.  Long-range excitons in the optical excitations
have been characterized, and the long-range component of oscillator
strengths is calculated as a function of the torsion angle.
We have shown that the torsion effects in PPP are relatively small. 
In contrast, the torsions of PAN decrease the long-range component 
by about half of the magnitudes from that of the torsion-free system.

\mbox{}

\begin{flushleft}
{\bf Acknowledgements}
\end{flushleft}

Useful discussion with Y. Shimoi, S. Abe, and K. Murata
is acknowledged.  Numerical calculations have been 
performed on the DEC AlphaServer of Research Information
Processing System Center (RIPS), Agency of Industrial 
Science and Technology (AIST), Japan.

\pagebreak
\begin{flushleft}
{\bf References}
\end{flushleft}

\noindent
$[1]$ K. Harigaya, J. Phys. Soc. Jpn. {\bf 66}, 
1272 (1997).\\
$[2]$ K. Harigaya, J. Phys.: Condens. Matter {\bf 9},
5253 (1997).\\
$[3]$ K. Harigaya, J. Phys.: Condens. Matter {\bf 9},
5989 (1997).\\
$[4]$ D. Chen, M. J. Winokur, M. A. Masse, and F. E. Karasz,
Phys. Rev. B {\bf 41}, 6759 (1990).\\
$[5]$ J. L. Baudour, Y. Delugeard, and P. Rivet,
Acta Crystallogr. B {\bf 34}, 625 (1978).\\
$[6]$ M. E. Jozefowicz, R. Laversanne, H. H. S. Javadi, 
A. J. Epstein, J. P. Pouget, X. Tang, and A. G. MacDiarmid,
Phys. Rev. B {\bf 39}, 12958 (1989).\\
$[7]$ Z. G. Soos, S. Etemad, D. S. Galv\~{a}o, and S. Ramasessha,
Chem. Phys. Lett. {\bf 194}, 341 (1992).\\
$[8]$ J. M. Ginder and A. J. Epstein, Phys. Rev. B
{\bf 41}, 10674 (1990).\\


\begin{flushleft}
{\bf Figure Captions}
\end{flushleft}

\mbox{}

\noindent
Fig. 1.  Polymer structures of (a) PPP and (b) PAN.
These figures are only schematic.  Geometries with ring 
torsions and the zigzag chain structure of PAN are used 
in the actual calculations.

\mbox{}

\noindent
Fig. 2. Optical absorption spectra of the PPP shown in arbitrary
units.  The polymer axis is parallel to the $x$-axis.  The polymer 
without ring torsions is in the $x$-$z$ plane.  The electric field 
of light is parallel to the chain and in the direction of the 
$x$-axis in (a), and it is perpendicular to the chain and is along 
with the $y$-axis in (b).  It is along with the $z$-axis
in (c).  The orientationally averaged spectra are shown in (d).
The number of the PPP monomer units is $N_{\rm m}= 20$.  
The bold line is for the total absorption.  The thin line 
indicates the absorption of the long-range component.  
The Lorentzian broadening $\gamma = 0.15 t$ is used.

\mbox{}

\noindent
Fig. 3.  Long-range component of the optical absorption
spectra as a function of the torsion angle in PPP.  The 
monomer unit number is $N_{\rm m}=20$.  The closed squares 
are for the orientationally averaged absorption.  The open 
circles, squares, and triangles indicate the data for the 
cases with the electric field parallel to the $x$-, $y$-,
and $z$-axes, respectively.

\mbox{}

\noindent
Fig. 4. Optical absorption spectra of the PAN shown in arbitrary
units.  The polymer axis is parallel to the $x$-axis.  The polymer 
without ring torsions is in the $x$-$z$ plane.  The electric field 
of light is parallel to the chain and in the direction of the 
$x$-axis in (a), and it is perpendicular to the chain and is along 
with the $y$-axis in (b).  It is along with the $z$-axis
in (c).  The orientationally averaged spectra are shown in (d).
The number of the PAN monomer units is $N_{\rm m}= 20$.  
The bold line is for the total absorption.  The thin line 
indicates the absorption of the long-range component.  
The Lorentzian broadening $\gamma = 0.15 t$ is used.

\mbox{}

\noindent
Fig. 5.  Long-range component of the optical absorption
spectra as a function of the torsion angle in PAN.  The 
monomer unit number is $N_{\rm m}=20$.  The closed squares 
are for the orientationally averaged absorption.  The open 
circles, squares, and triangles indicate the data for the 
cases with the electric field parallel to the $x$-, $y$-,
and $z$-axes, respectively.

\end{document}